\documentclass[12pt]{article}
\usepackage{geometry}
\newcommand{\beq}{\begin{equation}}
\newcommand{\eeq}{\end{equation}}
\newcommand{\beqs}{\begin{eqnarray}}
\newcommand{\eeqs}{\end{eqnarray}}

\newcommand{\Q}{{\cal Q}}

\begin{document}

\begin{titlepage}

\hfill ULB-TH/08-33

\vspace{40pt}

\begin{center}
{\Large \bf Boosting Taub-NUT to a BPS NUT-wave}

\end{center}

\vspace{15pt}

\begin{center}
{\large  Riccardo Argurio, Fran\c cois Dehouck and Laurent Houart}\\

\vskip 35pt

Physique Th\'eorique et Math\'ematique and International Solvay Institutes \\
Universit\'e Libre de Bruxelles, C.P. 231, 1050 Bruxelles, Belgium
\end{center}

\vspace{20pt}

\begin{center}
\textbf{Abstract}
\end{center}
 The boosted Taub-NUT metric with zero ADM
    mass is shown to possess a dual momentum in the direction of the
    boost giving credit to the existence of another
4-vector $K_{\mu}$ in linearized gravity associated to the NUT
    charge and dual to the usual $P_{\mu}$.
    Taking the infinite boost limit we obtain a shock pp-wave with NUT charge.  We show that the latter is the  gravitational dual  of the infinitely boosted Schwarzschild metric, also known as the
    Aichelburg-Sexl pp-wave. We review the fact that this new shock
    pp-wave is also an half-BPS solution of $\mathcal{N}=1$ supergravity. It has a BPS
    bound equal to $K_{t}=|K_{z}|$.

\end{titlepage}

\section{Introduction}
Electric-magnetic duality  was originally established as a
symmetry of the Maxwell equations and revealed itself a powerful
tool. As examples let us remind Dirac's realization \cite{Dirac:1931kp} that the very
existence of magnetic monopoles would imply the quantization of
charges, or Montonen and Olive \cite{Montonen:1977sn} who conjectured the presence of this
symmetry inside non-abelian gauge theories
which was later shown to hold  as a strong-weak duality in $\mathcal{N}=4$ super Yang-Mills  \cite{Osborn:1979tq}.

Although Einstein's equations of General Relativity and the
concept of electric-magnetic duality were already
well-established, it was only in the 50s-60s that Taub \cite{Taub
:1951tr}  and Newman-Unti-Tamburino \cite{Newman Tamburino Unti
:1963tr} discovered a solution, called the Taub-NUT solution,
which possesses a mass but also another parameter called the NUT
charge. It was soon realized \cite{Demianski Newman :1966tr} that
this NUT charge could be understood as the gravitational magnetic
dual of the ADM mass (see for example \cite{Bunster:2006rt} and
references therein). However, up to now, this
duality, which acts as an Hodge operator on the Riemann tensor,
has only been verified in linearized gravity and still resists
attempts to be
proven in the full non-linear theory \cite{Henneaux:2004jw,Deser:2005sz}.

Motivated by the understanding of this important duality that seems
to exist in General Relativity and the
already-known BPS bound $  M^2+N^2=Z^2$ \cite{Kallosh:1994ba}, we were first interested in the
supersymmetric properties of the charged Taub-NUT.
We reviewed in our previous paper \cite{Argurio Dehouck
Houart:2007tr} how to obtain this BPS bound in ${\cal N}=2$ supergravity,
and derived the Killing spinors for this particular solution. We
then showed the impossibility of including the NUT charge in the
usual supersymmetry algebra and proposed a way of modifying the
algebra such as to include it. This discussion was motivated
by the projection we obtained on the Killing spinors and from
Nester's construction. On the way, we also derived generalized
expressions for the ADM and dual ADM 4-momenta. This construction
made it clear that the vielbein formalism is certainly more
appropriate to study the duality as it permits to express surface
integrals in terms of regular spin connections (along the Misner string direction). It was also noted that this completion of the
$\mathcal{N}=2$
supersymmetry algebra should already be present in $\mathcal{N}=1$.

In this work we would like to give credit to the modification of
the $\mathcal{N}=2$ supersymmetry algebra presented in
\cite{Argurio Dehouck Houart:2007tr}.
We provide evidence that the modification
should also be
present in $\mathcal{N}=1$ and verify that it is indeed proportional to  a 4-vector
$K_{\mu}$, where $K_0=N$ for the Taub-NUT solution.
In this letter, we thus consider the Taub-NUT solution with the ADM
mass set to zero and with no electromagnetic charges. We shall
refer to this solution as the ``pure Taub-NUT". It is understood
that this metric is solution of the vacuum Einstein equations and
that it preserves no supersymmetries in $\mathcal{N}=1$
supergravity as the BPS bound is $N=0$.

The paper is organized as follows: In Section 2,  we apply our charge
formul\ae \cite{Argurio Dehouck Houart:2007tr} to the boosted pure
Taub-NUT and show that $K_{\mu}$ does transform as $P_{\mu}$ under
boosts. In Section 3, we study the limit of infinite boost
using the  method of Aichelburg-Sexl and show that we obtain a
shock pp-wave. As expected from the result in
\cite{Bunster:2006rt} where it is checked that linearized pure
Taub-NUT is dual to linearized Schwarzschild, we also recover the
infinitely boosted Taub-NUT metric as the gravitational dual of
the Aichelburg-Sexl pp-wave, or infinitely boosted
Schwarzschild \cite{Aichelburg Sexl:1971tr}. In Section 4, we
review the supersymmetric properties of pp-waves and show that the
dual pp-wave preserves half of the supersymmetries and satisfies the BPS
bound $K_0=-K_3$. This is also checked calculating
the charges of the dual pp-wave.

\section{The boosted Taub-NUT solution}

In this Section, we wish to add credit to the existence
of the 4-vector $K_\mu$. This 4-vector was already presented in the
derivation of the quantization condition in gravity in
\cite{Bunster:2006rt}. We also showed in \cite{Argurio Dehouck
Houart:2007tr} that the Nester construction or a formal definition
of a gravitational dual analogue of the energy-momentum
 $P_{\mu}$ gives a unique expression for the 4-vector $K_\mu$. The charge formul{\ae} were however only applied to the static Taub-NUT.
 To show that this $K_{\mu}$ does transform as a 4-vector, we will boost the Taub-NUT solution  and show that a momentum contribution appears
 in the boosted direction.

To do that, let us first recall the metric of  the Taub-NUT solution which
possesses a mass $M$ and a NUT charge $N$:
\begin{eqnarray}
ds^2 = -\frac{\lambda}{R^2}(dt+2N\cos\theta d\phi)^2+
\frac{R^2}{\lambda} dr^2 +R^2(d\theta^2+\sin^2\theta
d\phi^2)
\end{eqnarray}
where we defined $\lambda=r^2-N^2-2Mr$ and $R^2=r^2+N^2$. From now on, we will set $M=0$ and discuss the pure Taub-NUT solution.

As we showed in \cite{Argurio Dehouck Houart:2007tr}, if we work
in the vielbein formalism it is possible to derive an expression for $P_{\mu}$
and $K_{\mu}$ as surface integrals over the spin connection.  The
linearized vielbein is given by:
\begin{equation}
e^\mu=dx^\mu+\frac{1}{2}
\eta^{\mu\nu}(h_{\nu\rho}+v_{\nu\rho})dx^\rho
\end{equation}
where $h_{\nu\rho}=h_{\rho\nu}$ is the linearized metric,
$v_{\nu\rho}=-v_{\rho\nu}$ is related to local Lorentz invariance
and no difference is made between flat and curved indices as we
are working in linearized gravity around cartesian flat coordinates. The expressions for the charges
in terms of the vielbein were found to be \cite{Argurio Dehouck
Houart:2007tr}:
\begin{eqnarray}
P_0 & = & \frac{1}{16\pi} \oint ( \partial_i h_{li}-  \partial_l h_{ii}+ \partial_i v_{il}) d\hat \Sigma_l, \label{p0v}\\
P_k & = & \frac{1}{16\pi} \oint ( \partial_0 h_{lk}- \partial_l h_{0k}
+ \delta^k_l \partial_{i} h_{0i} - \delta^k_l \partial_{0} h_{ii} +\partial_{k} v_{0 l}+ \delta^k_l \partial_{i} v_{i 0} ) d\hat \Sigma_l, \label{pkv}\\
K_0 & = & \frac{1}{16\pi} \oint  \epsilon^{lij}  ( \partial_i h_{0j}+ \partial_j v_{i0} )  d\hat \Sigma_l , \label{k0v}\\
K_k & = & \frac{1}{16\pi} \oint  \epsilon^{lij}  ( \partial_i h_{kj}+ \partial_j v_{ik} )  d\hat \Sigma_l , \label{kkv}
\end{eqnarray}
where  $\epsilon^{123}=1$. The only restriction for using these
expressions is that the spin connection, using a particular
linearized vielbein, has to be regular. For the Taub-NUT metric
we obtained $K_0=N$, $P_0=M$ and $P_i=K_i=0$ using a triangular vielbein.

One could easily argue that there always exists a gauge
transformation such that the vielbein can be set in a symmetric
gauge (and by this we mean $v_{\mu\nu}=0$) which would reduce our
formul{\ae}  for $P_\mu$ to the standard ADM ones. However, one should
be aware that this is only valid (at the level of the calculation
of charges) in the case where the gauge transformation is
non-singular. In other words, one can use
these formul{\ae} in the symmetric gauge where $v_{\mu\nu}=0$ only if
the spin connection is regular along the Misner string direction. As stated in \cite{Argurio Dehouck
Houart:2007tr} we see that these expressions are generalized by
saying that we should not fix the gauge in
the symmetric gauge but rather in the ``regular spin-connection" gauge.

A first natural test to certify the existence of the 4-vector
$K_{\mu}$ is to show that the boosted Taub-NUT has
$K_0=\gamma N$ and a momentum in the direction of the boost equal to
$\gamma \beta N$. As we are interested in calculating a surface
integral at spatial infinity, we will directly work with the
linearized pure Taub-NUT:
\begin{eqnarray}
ds^2_{Lin}= - d\bar{t}^2 - 4 N \cos\bar{\theta}
d\bar{\phi} d\bar{t}+d\bar{r}^2+\bar{r}^2(d\bar{\theta}^2+\sin^2\bar{ \theta} d{\bar{\phi}}^2)
\end{eqnarray}
which can be written in cartesian coordinates as:
\begin{eqnarray}
\label{taublin}
ds^2_{Lin}= - d\bar{t}^2 - 4N \frac{\bar{z}}{\bar{r}}\: \frac{1}{\rho^2}
(\bar{x}d\bar{y}-\bar{y}d\bar{x})d\bar{t}+d\bar{x}^2+d\bar{y}^2+d\bar{z}^2
\end{eqnarray}
where $\rho^2=\bar{x}^2+\bar{y}^2$.

If we now perform a boost in the $z$-direction:
\begin{eqnarray}
\label{boostz}
\bar{t}&=& \gamma(t-\beta z)\: \:\:\:\:\:
\bar{z}= \gamma(z-\beta t)\nonumber \\
\bar{x}&=& x \:\:\:\:\:\:\:\:\:\:\:\:\:\:\:\:\:\:\:\:\:\: \bar{y}=y
\end{eqnarray}
we get:
\begin{eqnarray}
\label{boostedmetric2}
ds^2&=& -dt^2 +dx^2+dy^2+ dz^2 -
4N \frac{\gamma^2(z-\beta t)}{\bar{r}\rho^2}
(dt-\beta dz)(xdy-ydx)
\end{eqnarray}

One should be careful while treating the coordinate $\bar{r}$, as
the large radius limit is really $ r \equiv(x^2+
y^2 +z^2)^{1/2} \rightarrow \infty$, and thus:
\begin{eqnarray}
 \frac{1}{\bar{r}} &\equiv&  \biggr [ x^2+ y^2+ \gamma^2(z-\beta \: t)^2 \biggl]^{-1/2}\nonumber \\
   &=& \frac{1}{r} \biggr [(\sin^2\theta +\gamma^2 \cos^2\theta) +\gamma^2 \beta^2 (t^2/ r^2)-2\gamma^2 \beta \cos\theta
   (t/r) \biggl ]^{-1/2} \nonumber \\
 & \sim& \frac{1}{rB}
+O(1/r^2)
\end{eqnarray}
where we defined $B=\sqrt{sin^2\theta +\gamma^2 cos^2\theta}$.

Our choice for the vielbein is:
\begin{eqnarray}\label{triagviel}
e^0&=& dt- 2N \frac{\gamma^2 (z-\beta t)}{\bar{r}\rho^{2}}(ydx-xdy)  \nonumber \\
e^1&=&  dx \nonumber \\
e^2&=&  dy \nonumber \\
e^3&=& - 2N \frac{\gamma^2 \beta (z-\beta
t)}{\bar{r}\rho^{2}}  (y \: dx-
x \: dy) +  dz
\end{eqnarray}
where it can be checked that the spin connection is regular, in
agreement with \cite{Argurio Dehouck Houart:2007tr} because our
choice is precisely the triangular vielbein $e^{\bar{m}}_{\:{\bar \mu}}$ for
the linearized static Taub-NUT metric transformed under the boost
to $e^{m}_{\: \mu}=\Lambda^{m}_{\:\:\:\bar{n}}
\:\Lambda_{\mu}^{\:\:\: \bar{\nu}} e^{\bar{n}}_{\: \bar{\nu}}=
\delta^{m}_{\: \mu}+ \frac{1}{2} \eta^{m \nu}(h_{{\nu}{\mu}}+v_{{\nu}
{\mu}})$.

 Looking at (\ref{boostedmetric2}), the linear
perturbations are:
\begin{eqnarray}
h_{xz}&=& -\beta h_{tx}=  -2 N  \gamma^2 \beta \frac{(z-\beta t)}{\bar{r}} \: \frac{y}{x^2+y^2}\nonumber \\
h_{yz}&=&  -\beta h_{ty}=\:\: 2 N
\gamma^2 \beta \frac{(z-\beta t)}{\bar{r}} \:
\frac{x}{x^2+y^2}
\end{eqnarray}
And we can directly see in (\ref{triagviel}) that our vielbein gives us $v_{ta}= h_{ta} $ and $v_{za}= h_{za}$ for $a=x,y$.

We can now easily proceed to the calculation of $K_0$:
\begin{eqnarray}\label{K0}
K_{0}&=&  \frac{1}{16\pi} \oint  \epsilon^{lij}  ( \partial_i h_{0j}+ \partial_j v_{i0} )  d\hat \Sigma_l
= \frac{1}{8\pi} \oint  \epsilon^{lij}  \partial_i h_{0j}   d\hat \Sigma_l \nonumber\\
&=& \frac{N}{4\pi}\gamma^2 \oint_{S} \frac{\sin\theta}{B^3} d\theta d\phi
\nonumber \\
&=& \gamma N.
\end{eqnarray}
Note also that the time dependence in the integrand (\ref{K0}) is subleading and tends to zero when $r \rightarrow \infty$.

The calculation for $K_z$ is readily the same and we find:
\begin{eqnarray}\label{Kz}
K_z &=&- \frac{\beta}{8\pi} \oint  \epsilon^{lij}  \partial_i h_{0j}  d\hat \Sigma_l = -\beta K_0=-\gamma \beta N
\end{eqnarray}
while $K_x=K_y=0$. Finally using (\ref{p0v}) and (\ref{pkv}), it is not difficult to show that
  $P_{\mu}=0$ for the boosted Taub-NUT solution.\footnote{For $P_0$ and $P_z$ the integrands  are zero while for $P_x$ and $P_y$ the integrands are non vanishing but the integrals  (\ref{pkv}) are zero.}

We have thus shown that $K_{\mu}$ behaves as a 4-vector.

\section{The pp-wave and its magnetic dual}

In this section, we present two ways of obtaining the infinite
boost of the Taub-NUT metric. The first derivation follows the
steps of the method of Aichelburg and Sexl \cite{Aichelburg
Sexl:1971tr} who performed the infinite boost of the Schwarzschild
metric. They obtained a (shock) pp-wave
given by the expression:
\begin{eqnarray}
\label{axse}
ds^2= -dt^2+dx^2+dy^2+dz^2-8 \: p\: \ln(\sqrt{x^2+y^2})\:
\delta(t-z) \: (dt-dz)^2
\end{eqnarray}
This method was generalized  in
\cite{Lousto Sanchez:1989tr} and used, for example, for the
infinite boost of the Reissner-Nordstr\"om black hole. This more
general analysis was used in \cite{Hayashi Samura:1994tr} for the
infinite boost of  the Kerr black holes where
the  Aichelburg-Sexl metric is
shown to be recovered in a certain limit.

 One important characteristic of (\ref{axse})
is that this metric is solution of the linearized but also of the
full Einstein's equations. In fact, this kind of solutions was
already well-known. The Aichelburg-Sexl  metric belongs to the
wider class of pp-waves, plane fronted waves with parallel rays,
first introduced by Brinkmann in 1925 as metrics on Lorentzian
manifolds:
\begin{eqnarray}
\label{fullform}
ds^2= H(u,x,y) du^2 - du \: dv +dx^2+dy^2
\end{eqnarray}
where $H$ is a smooth function. If moreover the function $H$ is
harmonic in $x$ and $y$
then it is a solution of the  full Einstein's equations.

The specificity of the Aichelburg-Sexl solution is that the $H$ function
factorizes its $u$ dependence in a delta function such that
$H(u,x,y)=F(x,y) \delta(u)$ and $F(x,y)$ is a harmonic function.
This shock pp-wave was also discussed in  \cite{'t Hooft: 1985}
and understood as the gravitational radiation of a particle
travelling at the velocity
of light measured by an observer at rest.

We now perform the infinite boost on the Taub-NUT solution.
Like in the Schwarzschild case \cite{Aichelburg
Sexl:1971tr}, we only need the linearized part
of the metric. For the pure Taub-NUT we thus have:
\begin{eqnarray}
ds^2= -d\bar{t}^2 +d\bar{x}^2+d\bar{y}^2+ d\bar{z}^2+ds_{def}^2
\end{eqnarray}
where $ds_{def}^2=-4N \cos\bar{\theta} d\bar{t}\: d\bar{\phi}$.

Here, for convenience, we will take the Misner string along the $x$ direction (namely interchanging $\bar{x}$ and $\bar{z}$ in (\ref{taublin})) and boost along the
$z$ direction according to (\ref{boostz}).
We then find in the leading order in
$\gamma$:
\begin{eqnarray}
\bar{t}& \rightarrow & \gamma \: u \nonumber \\
\bar{z}& \rightarrow & -\gamma \: u \nonumber \\
\bar{r}^2 & \rightarrow & \gamma^2 \: u^2+ \: (x^2+y^2) \nonumber \\
\cos\bar{\theta}=\frac{\bar{x}}{\bar{r}}& \rightarrow & x/ \sqrt{\gamma^2 \: u^2+(x^2+y^2)} \nonumber\\
d\bar{\phi} & \rightarrow & \frac{1}{\gamma} \: \frac{y du }{u^2+
\gamma^{-2} y^2}
\end{eqnarray}
where $\tan\bar{\phi}= \bar{y}/ \bar{z}$,  we defined $u=t-\beta z$, and by leading order we mean that
$\bar{z}\rightarrow \gamma (z-\beta t)= -\gamma u +\gamma
(1-\beta) (z+t)\sim -\gamma u$. Note that we can also drop in $d\bar{\phi}$  the
second term in $u \: dy$ as we will see
that it is at $u=0$,
when the infinite boost limit is considered, that a contribution appears.

The deformed part of the metric becomes:
\begin{eqnarray}\label{a}
ds_{def}^2&=&-4N  \frac{x}{ \sqrt{\gamma^2 \: u^2+(x^2+y^2)}}\:
\gamma du \: \: \frac{1}{\gamma} \: \frac{y du
}{u^2+\gamma^{-2}y^2}
\end{eqnarray}
In the limit of infinite boost, we take $\gamma\rightarrow \infty
$ and $N \rightarrow 0$ while keeping $N\gamma=k$. This means we
have:
\begin{eqnarray}\label{b}
ds^2_{def}= -8k \:  \lim_{\epsilon\rightarrow 0} \:  \frac{1}{\epsilon}
\:\: \frac{A \, du^2}{2\sqrt{(u/\epsilon)^2+(1+A^2)}((u/\epsilon)^2+A^2)}
\end{eqnarray}
where we wrote $\epsilon= \gamma^{-1}x$ and $A=y/x$.

If we now take $\epsilon\rightarrow 0$ in the sense of the
distributions using the fact that:
\begin{eqnarray}\label{distrib}
\lim_{\epsilon\rightarrow 0} \frac{1}{\epsilon}
f(z/\epsilon)=\delta(z)
\end{eqnarray}
for a function $f$ such that
$\int^{+\infty}_{-\infty} f(z) dz=1,$
we find:
\begin{eqnarray}
ds_{def}^2= -8k \: \arctan(1/A) \: \delta(u)\: du^2= -8k \: \arctan
(x/y) \:  \delta(u) \: du^2
\end{eqnarray}
and the metric of the infinitely boosted pure Taub-NUT metric is:
\begin{eqnarray}\label{NUTtywave}
ds^2=-dt^2+dx^2+dy^2+dz^2 -8k \: \arctan(x/y) \: \delta(t-z) \:
(dt-dz)^2
\end{eqnarray}
The metric (\ref{NUTtywave}) is obviously solution of the full non-linear Einstein equations because it is of the form
(\ref{fullform}) and $\arctan(x/y)$ is harmonic.
Moreover, we show in the next Section that $K_0=-K_z=k$ as it should.

To confirm this limit we now describe an alternative derivation using the gravitational duality of (linearized) gravity
\cite{Bunster:2006rt}. We show that the infinite boost of the
Taub-NUT metric is  the gravitational dual of the Aichelburg-Sexl
pp-wave. In fact it is enough to check that one metric has a
Riemann tensor dual to the other. For simplicity, we will not take
into account the Dirac delta function as it does not affect the following
analysis.

The non-trivial fluctuations for the Aichelburg-Sexl pp-wave are:
\begin{eqnarray}
h_{tt}=h_{zz}=-h_{tz}= -8\: p\: \ln(\sqrt{x^2+y^2})
\end{eqnarray}

The linearized Riemann tensor is defined as:
\begin{eqnarray}
R_{\alpha \beta \gamma \delta}= 2 \partial_{[\alpha} h_{\beta
][\gamma,\delta]}
\end{eqnarray}
whose non-trivial components for the Aichelburg-Sexl metric are:
\begin{eqnarray}\label{Riem}
R_{tatb}= -\frac{1}{2} \: \partial_a \partial_b h_{tt} \:\:\:\:\:
R_{tazb}= -\frac{1}{2} \: \partial_a \partial_b h_{tz} \:\:\:\:\:
R_{zazb}= -\frac{1}{2} \: \partial_a \partial_b h_{zz}
\end{eqnarray}
for $a,b=x,y$ and where $R_{\alpha\beta\gamma\delta}$ is, as usual, antisymmetric in its first
two and last two indices and symmetric under the exchange of the first
and second pair
of indices.

The infinitely boosted pure Taub-NUT has non-trivial fluctuations:
\begin{eqnarray}
\tilde{h}_{tt}=\tilde{h}_{zz}=-\tilde{h}_{tz}= -8\: k\:
\arctan(x/y)
\end{eqnarray}
where $\tilde{h}_{\mu\nu}$ refers to the dual metric. The non-trivial
components of the Riemann tensor are thus
the same as in (\ref{Riem}) but with $h_{\mu\nu}$ replaced by $\tilde{h}_{\mu\nu}$.

It is then easy to check  that the non-trivial components
of the Riemann tensor for (\ref{NUTtywave}) are precisely the
ones obtained from (\ref{Riem}) by duality using:
\begin{eqnarray}
\tilde{R}_{\alpha \beta \lambda \mu}=\frac{1}{2} \: \epsilon_{\alpha \beta
\gamma \delta} \: R^{\gamma \delta}_{\:\:\:\: \lambda \mu}
\end{eqnarray}
with $\epsilon_{txyz}=1$.

We have thus checked  that the dual of the Aichelburg-Sexl pp-wave
is the infinitely boosted Taub-NUT, which we will call the dual
pp-wave or NUT-wave. The metric is another shock pp-wave but with a different
harmonic function:
\begin{eqnarray}
ds^2= -dt^2+dx^2+dy^2+dz^2-8 \: k\: \arctan(x/y)\: (dt-dz)^2
\end{eqnarray}

Note that the harmonic function has a
cut in the $x-y$ plane, remnant of the Misner string singularity.
It is interesting to recall \cite{Argurio Dehouck Houart:2007tr} that in the
Killing spinor equation for Taub-NUT,  the ADM mass and NUT charge
appear in the combination $M- \gamma_5 N$ where $\gamma_5^2= -1$, which is reminiscent of a complex structure. In the same way, for the pp-waves, we could construct a
complex variable $\zeta=y+ix$  whose logarithm is $\ln \zeta= \ln
\sqrt{x^2+y^2} + i \arctan(x/y)$ and attribute the real part
of this logarithm to the Aichelburg-Sexl metric and the imaginary part
to the dual pp-wave. This last fact can be generalized to any solution (\ref{fullform}), where $H(u,x,y)=F(x,y) \delta(u)$ and $F(x,y)$ is a harmonic function. The gravitational dual solution is characterized by $H(u,x,y)=\tilde{F}(x,y) \delta(u)$ where $\tilde{F}$ is the harmonic conjugate function of $F$ (namely ${\cal F}(\zeta)= F+i \tilde{F}$ is an holomorphic function of $\zeta$).\footnote{The holomorphic nature of ${\cal F}(\zeta)$ is reminiscent of the holomorphic nature of the complex Ernst potential for BPS solutions (see for instance Section 3.4 of \cite{Englert:2007qb}).}

\section{Charges and supersymmetric properties of the dual pp-wave}

In this section, we want to review the fact that the shock pp-wave
is a supersymmetric solution of $\mathcal{N}=1$
supergravity\footnote{Note that all supersymmetric solutions of
$\mathcal{N}=1$ supergravity were classified in \cite{Tod:1983pm}.} and, as the
BPS bound is $P_0=-P_3$ for the Aichelburg-Sexl metric, we want
to establish that the BPS bound is $K_0=-K_3$ for our dual
pp-wave. As a
final check, we show that the charges for the
dual pp-wave verify this BPS bound as it can be
expected from the infinite boost of (\ref{K0}) and (\ref{Kz}).

To use our formulae for the charges, we need a regular spin
connection \cite{Argurio Dehouck Houart:2007tr}. We will give arguments that the good choice of
vielbein is the symmetric one. To do that, let us start with a
 pp-wave of the form:
\begin{eqnarray}
ds^2&=& -dt^2+dx^2+dy^2+dz^2+ F (dt-dz)^2
\nonumber \\
 &=& -du(dv-F du)+ dx^2 + dy^2
\end{eqnarray}
where $F=F(x,y)$ and where we defined light-cone coordinates
$u=t-z$ and
$v=t+z$. Note that we dropped again the delta function for simplicity.

An obvious vielbein choice in light-cone coordinates is:
\begin{eqnarray}
e^-&=& du\:\:\:\:\:\:\:\:\: e^+= dv-F du\nonumber \\
e^1&=& dx\:\:\:\:\:\:\:\:\: e^2= dy
\end{eqnarray}
and the metric is $ds^2=\eta_{ab} e^a \: e^b$ where
the non-vanishing components are $\eta_{11}=\eta_{22}=1$,
$\eta_{+-}=\eta_{-+}=-1/2$.

Going back to cartesian coordinates, we obtain the symmetric
vielbein:
\begin{eqnarray}
e^0&=& \frac{1}{2}(e^+ + e^-)= dt-\frac{F}{2} (dt-dz)
 \nonumber \\
e^1&=& dx \nonumber \\
 e^2&=& dy \nonumber \\
 e^3&=& \frac{1}{2}(e^+ - e^-)=dz -\frac{F}{2}(dt-dz)
\end{eqnarray}
where symmetricity is understood by the fact that
$v_{\mu\nu}=-v_{\nu \mu}=0$. The non-trivial components of the spin
connection are:
\begin{equation}
\omega_{0a}=-\omega_{3a}= \frac{1}{2} \partial_a F(x,y)(dt-dz)
\end{equation}
where $F(x,y)=-8\: k \:  \arctan(x/y)$ for the dual pp-wave. Even
if in the case of our dual pp-wave the metric has a string
singularity, one can see that the spin connection is ``regular" in
the $x-y$ plane . One could argue that a triangular vielbein with
a regular linearized spin connection could also be used. However,
it is important to note that our choice of vielbein is
linear in the full theory. If one tries to construct such a triangular vielbein for
example it would not be linear in the full theory and the
operation of linearizing would then erase singularities in the spin connection such as
$1/ \sqrt{1-F} \sim 1+ (1/2) F$ with $F$ being the singular
harmonic function. The calculation of charges would then fail.

It can be easily seen that the pp-wave solution is an half-BPS
solution of $\mathcal{N}=1$ supergravity when looking at the
Killing spinor equation (conventions are taken from \cite{Argurio
Dehouck Houart:2007tr}):
\begin{eqnarray}
\delta \psi_{\mu}= \biggr [ \partial_{\mu}+\frac{1}{4}
\omega_{\mu}^{mn}\: \gamma_{mn} \biggl ] \epsilon=0
\end{eqnarray}

This gives us the set of equations:
\begin{eqnarray}
\delta \psi_{t}&=& \biggr [ \partial_{t}- \frac{1}{4}
\partial_a F(x,y) (\gamma_{0}+\gamma_{3})\gamma_a \biggl ] \epsilon=0 \nonumber \\
\delta \psi_{x}&=& \partial_{x} \epsilon =0 \nonumber \\
\delta \psi_{y}&=& \partial_{y} \epsilon =0 \nonumber \\
\delta \psi_{z}&=& \biggr [ \partial_{z}+\frac{1}{4}
\partial_a F(x,y) (\gamma_{0}+\gamma_{3})\gamma_a \biggl ] \epsilon=0
\end{eqnarray}
As the second and third equations show that $\epsilon$ does not
depend on $x$ and $y$, then the first and fourth equations imply
the projection $(\gamma_0+\gamma_3)\epsilon=0$. This determines that the
solution preserves half of the supersymmetries and has a constant
Killing spinor. This projection corresponds to the BPS bound
$K_0=-K_3$
for our dual pp-wave.

As a final check, we calculate the charges for the dual pp-wave.
For the Aichelburg-Sexl pp-wave, this was done in \cite{Aichelburg
Balasin:1998tr}. Here, in the symmetric vielbein, we have $v_{\mu
\nu}=0$ such that:
\begin{eqnarray}
K_0 & = & \frac{1}{16\pi} \oint  \epsilon^{lij} \partial_i h_{0j}  d\hat \Sigma_l= -\frac{k}{2\pi} \oint \frac{1}{r} \: \delta(t-z) \:   d\hat \Sigma_r \nonumber \\
    & = & k \oint r \: \sin\theta \:  \delta(t-r\cos\theta) \:   d\theta= k \oint  \: \delta(t-r\cos\theta) \:   d(r \cos \theta) \nonumber\\
    &=& k
\end{eqnarray}
Again the calculation for $K_z$ is readily the same and gives $-k$.
There is no contribution to $P_{\mu}$.
\section{Conclusions}
In this letter, we have provided some more arguments to the fact
that General Relativity, at least at the linear level, should include a 4-vector $K_{\mu}$ dual to the usual one $P_{\mu}$.

Moreover, we showed that the infinite boost of Taub-NUT is a shock
pp-wave and thus also an half-supersymmetric solution of
$\mathcal{N}=1$ supergravity. This provides more evidence
that the NUT charge should be included in the $\mathcal{N}=1$
supersymmetry algebra such as conjectured
in \cite{Argurio Dehouck Houart:2007tr}:
\begin{equation}
\{\Q, {\Q'}\} = \gamma^\mu C P_\mu +\gamma_5 \gamma^\mu C K_\mu
\label{superalgnutnew}
\end{equation}
where ${\Q'}$ is related to $\Q$ by a phase ${\Q'} = \Q \, e^{\alpha \gamma_5}$ with $\tan \alpha= K_0/P_0$.
Indeed, the ``modified" superalgebra (\ref{superalgnutnew}) is consistent with the projection and the BPS bound derived in the previous Section.

As a final word it would be interesting to see if the construction
of more general dual supersymmetric solutions with NUT charge
also provides modifications in their corresponding supersymmetric
algebras such as in the super-AdS algebra. It would also be interesting to study further the appearance of the Lorentzian Taub-NUT charges in  higher dimensional
supersymmetry algebras such as the one characterizing M-theory \cite{Hull:1997kt}.

\subsection*{Acknowledgments}
We would like to thank G.~Barnich, C.~Troessaert and V.~Wens for useful
discussions. This work was supported in part by
IISN-Belgium (conventions 4.4511.06, 4.4505.86 and 4.4514.08) by
the European Commission FP6 RTN programme MRTN-CT-2004-005104, and
by the Belgian Federal Science Policy Office through the
Interuniversity Attraction Pole P5/27. R.A. and L.H. are Research
Associates of the Fonds de la Recherche Scientifique--F.N.R.S.
(Belgium).

\end{document}